\documentclass[%
 reprint,
superscriptaddress,
nofootinbib,
 amsmath,amssymb,
 aps,
floatfix,
longbibliography
]{revtex4-1}

\usepackage{lineno}
\usepackage{color}
\usepackage{graphicx}% Include figure files
\usepackage{dcolumn}% Align table columns on decimal point
\usepackage{bm}% bold math
\usepackage{multirow}
\def\rctn{\ensuremath {\vec{\gamma} n \rightarrow K^+\Sigma^-}}

\def\bsa{\ensuremath {\boldsymbol{\Sigma}}}

\begin{document}

% \preprint{APS/123-QED}

\title{Beam-spin asymmetry \bsa{} for $\Sigma^-$ hyperon photoproduction off the neutron}% Force line breaks with \\
%\thanks{A footnote to the article title}%
%Insert Institutions here

\newcommand*{\ANL}{Argonne National Laboratory, Argonne, Illinois 60439}
\newcommand*{\ANLindex}{1}
\affiliation{\ANL}
\newcommand*{\ASU}{Arizona State University, Tempe, Arizona 85287-1504}
\newcommand*{\ASUindex}{2}
\affiliation{\ASU}
\newcommand*{\CSUDH}{California State University, Dominguez Hills, Carson, California 90747}
\newcommand*{\CSUDHindex}{2}
\affiliation{\CSUDH}
\newcommand*{\CANISIUS}{Canisius College, Buffalo, NY}
\newcommand*{\CANISIUSindex}{3}
\affiliation{\CANISIUS}
\newcommand*{\CMU}{Carnegie Mellon University, Pittsburgh, Pennsylvania 15213}
\newcommand*{\CMUindex}{4}
\affiliation{\CMU}
\newcommand*{\CUA}{Catholic University of America, Washington, D.C. 20064}
\newcommand*{\CUAindex}{5}
\affiliation{\CUA}
\newcommand*{\SACLAY}{IRFU, CEA, Universit\'{e} Paris-Saclay, F-91191 Gif-sur-Yvette, France}
\newcommand*{\SACLAYindex}{6}
\affiliation{\SACLAY}
\newcommand*{\CFNS}{CFNS Stony Brook University, Stony Brook, New York 11794}
\newcommand*{\CFNSindex}{7}
\affiliation{\CFNS}
\newcommand*{\CNU}{Christopher Newport University, Newport News, Virginia 23606}
\newcommand*{\CNUindex}{7}
\affiliation{\CNU}
\newcommand*{\DUKE}{Duke University, Durham, North Carolina 27708-0305}
\newcommand*{\DUKEindex}{9}
\affiliation{\DUKE}
\newcommand*{\DUQUESNE}{Duquesne University, 600 Forbes Avenue, Pittsburgh, Pennsylvania 15282 }
\newcommand*{\DUQUESNEindex}{10}
\affiliation{\DUQUESNE}
\newcommand*{\FU}{Fairfield University, Fairfield, Connecticut 06824}
\newcommand*{\FUindex}{11}
\affiliation{\FU}
\newcommand*{\FIU}{Florida International University, Miami, Florida 33199}
\newcommand*{\FIUindex}{13}
\affiliation{\FIU}
\newcommand*{\FSU}{Florida State University, Tallahassee, Florida 32306}
\newcommand*{\FSUindex}{14}
\affiliation{\FSU}
\newcommand*{\GWUI}{The George Washington University, Washington, DC 20052}
\newcommand*{\GWUIindex}{15}
\affiliation{\GWUI}
\newcommand*{\HELMBONN}{Helmholtz-Institut fuer Strahlen- und Kernphysik, Universit\"{a}t Bonn, 53115 Bonn, Germany}
\newcommand*{\HELMBONNindex}{16}
\affiliation{\HELMBONN}

\newcommand*{\INFNCA}{INFN, Sezione di Catania, 95123 Catania, Italy}
\newcommand*{\INFNCAindex}{16}
\affiliation{\INFNCA}

\newcommand*{\INFNFE}{INFN, Sezione di Ferrara, 44100 Ferrara, Italy}
\newcommand*{\INFNFEindex}{16}
\affiliation{\INFNFE}
\newcommand*{\INFNFR}{INFN, Laboratori Nazionali di Frascati, 00044 Frascati, Italy}
\newcommand*{\INFNFRindex}{17}
\affiliation{\INFNFR}
\newcommand*{\INFNGE}{INFN, Sezione di Genova, 16146 Genova, Italy}
\newcommand*{\INFNGEindex}{18}
\affiliation{\INFNGE}
\newcommand*{\INFNRO}{INFN, Sezione di Roma Tor Vergata, 00133 Rome, Italy}
\newcommand*{\INFNROindex}{19}
\affiliation{\INFNRO}
\newcommand*{\INFNTUR}{INFN, Sezione di Torino, 10125 Torino, Italy}
\newcommand*{\INFNTURindex}{20}
\affiliation{\INFNTUR}
\newcommand*{\INFNPAV}{INFN, Sezione di Pavia, 27100 Pavia, Italy}
\newcommand*{\INFNPAVindex}{21}
\affiliation{\INFNPAV}
\newcommand*{\ORSAY}{Universit'{e} Paris-Saclay, CNRS/IN2P3, IJCLab, 91405 Orsay, France}
\newcommand*{\ORSAYindex}{22}
\affiliation{\ORSAY}
\newcommand*{\Juelich}{Institute fur Kernphysik (Juelich), Juelich, Germany}
\newcommand*{\Juelichindex}{23}
\affiliation{\Juelich}
\newcommand*{\JMU}{James Madison University, Harrisonburg, Virginia 22807}
\newcommand*{\JMUindex}{24}
\affiliation{\JMU}
\newcommand*{\KNU}{Kyungpook National University, Daegu 41566, Republic of Korea}
\newcommand*{\KNUindex}{25}
\affiliation{\KNU}
\newcommand*{\LAMAR}{Lamar University, 4400 MLK Blvd, PO Box 10046, Beaumont, Texas 77710}
\newcommand*{\LAMARindex}{26}
\affiliation{\LAMAR}
\newcommand*{\MISS}{Mississippi State University, Mississippi State, MS 39762-5167}
\newcommand*{\MISSindex}{27}
\affiliation{\MISS}
\newcommand*{\ITEP}{National Research Centre Kurchatov Institute - ITEP, Moscow, 117259, Russia}
\newcommand*{\ITEPindex}{28}
\affiliation{\ITEP}
\newcommand*{\NPICA}{Nuclear Physics Institute of the Czech Academy of Sciences, 250 68 \v{R}e\v{z}, Czechia} 
\newcommand*{\NPICAindex}{29}
\affiliation{\NPICA}
\newcommand*{\UNH}{University of New Hampshire, Durham, New Hampshire 03824-3568}
\newcommand*{\UNHindex}{29}
\affiliation{\UNH}
\newcommand*{\NMSU}{New Mexico State University, PO Box 30001, Las Cruces, New Mexico 88003, USA}
\newcommand*{\NMSUindex}{30}
\affiliation{\NMSU}
\newcommand*{\NSU}{Norfolk State University, Norfolk, Virginia 23504}
\newcommand*{\NSUindex}{31}
\affiliation{\NSU}
\newcommand*{\OHIOU}{Ohio University, Athens, Ohio  45701}
\newcommand*{\OHIOUindex}{32}
\affiliation{\OHIOU}
\newcommand*{\ODU}{Old Dominion University, Norfolk, Virginia 23529}
\newcommand*{\ODUindex}{33}
\affiliation{\ODU}
\newcommand*{\JLUGiessen}{II Physikalisches Institut der Universitaet Giessen, 35392 Giessen, Germany}
\newcommand*{\JLUGiessenindex}{34}
\affiliation{\JLUGiessen}
\newcommand*{\RPI}{Rensselaer Polytechnic Institute, Troy, New York 12180-3590}
\newcommand*{\RPIindex}{35}
\affiliation{\RPI}
\newcommand*{\MSU}{Skobeltsyn Institute of Nuclear Physics, Lomonosov Moscow State University, 119234 Moscow, Russia}
\newcommand*{\MSUindex}{37}
\affiliation{\MSU}
\newcommand*{\TEMPLE}{Temple University,  Philadelphia, Pennsylvania 19122 }
\newcommand*{\TEMPLEindex}{39}
\affiliation{\TEMPLE}
\newcommand*{\JLAB}{Thomas Jefferson National Accelerator Facility, Newport News, Virginia 23606}
\newcommand*{\JLABindex}{40}
\affiliation{\JLAB}

\newcommand*{\UNIONCOLL}{Union College, Schenectady, New York, 12308}
\newcommand*{\UNIONCOLLindex}{8}
\affiliation{\UNIONCOLL}

\newcommand*{\UCONN}{University of Connecticut, Storrs, Connecticut 06269}
\newcommand*{\UCONNindex}{8}
\affiliation{\UCONN}
\newcommand*{\UDFJC}{Universidad Distrital Francisco José de Caldas, Bogot\'{a}, Colombia}
\newcommand*{\UDFJCindex}{41}
\affiliation{\UDFJC}
\newcommand*{\UTFSM}{Universidad T\'{e}cnica Federico Santa Mar\'{i}a, Casilla 110-V Valpara\'{i}so, Chile}
\newcommand*{\UTFSMindex}{41}
\affiliation{\UTFSM}
\newcommand*{\FERRARAU}{Universita' di Ferrara , 44121 Ferrara, Italy}
\newcommand*{\FERRARAUindex}{12}
\affiliation{\FERRARAU}
\newcommand*{\BRESCIA}{Universit\`{a} degli Studi di Brescia, 25123 Brescia, Italy}
\newcommand*{\BRESCIAindex}{43}
\affiliation{\BRESCIA}
\newcommand*{\INSUBRIA}{Universit\`{a} degli Studi dell'Insubria, 22100 Como, Italy}
\newcommand*{\INSUBRIAindex}{42}
\affiliation{\INSUBRIA}

\newcommand*{\MESSINA}{Universit\`{a} degli Studi di Messina, 98166 Messina, Italy}
\newcommand*{\MESSINAindex}{42}
\affiliation{\MESSINA}

\newcommand*{\ROMAII}{Universit\`{a} di Roma Tor Vergata, 00133 Rome, Italy}
\newcommand*{\ROMAIIindex}{36}
\affiliation{\ROMAII}

\newcommand*{\UOGABES}{University of Gabes, 6072-Gabes, Tunisia}
\newcommand*{\UOGABESIndex}{44}
\affiliation{\UOGABES}

\newcommand*{\GLASGOW}{University of Glasgow, Glasgow G12 8QQ, United Kingdom}
\newcommand*{\GLASGOWindex}{44}
\affiliation{\GLASGOW}
\newcommand*{\URICH}{University of Richmond, Richmond, Virginia 23173}
\newcommand*{\URICHindex}{34}
\affiliation{\URICH}
\newcommand*{\SCAROLINA}{University of South Carolina, Columbia, South Carolina 29208}
\newcommand*{\SCAROLINAindex}{38}
\affiliation{\SCAROLINA}
\newcommand*{\YORK}{University of York, York YO10 5DD, United Kingdom}
\newcommand*{\YORKindex}{45}
\affiliation{\YORK}
\newcommand*{\VT}{Virginia Tech, Blacksburg, Virginia 24061-0435}
\newcommand*{\VTindex}{46}
\affiliation{\VT}
\newcommand*{\VIRGINIA}{University of Virginia, Charlottesville, Virginia 22901}
\newcommand*{\VIRGINIAindex}{47}
\affiliation{\VIRGINIA}
\newcommand*{\WM}{College of William and Mary, Williamsburg, Virginia 23187-8795}
\newcommand*{\WMindex}{48}
\affiliation{\WM}
\newcommand*{\YEREVAN}{Yerevan Physics Institute, 375036 Yerevan, Armenia}
\newcommand*{\YEREVANindex}{49}
\affiliation{\YEREVAN}

\newcommand*{\NOWOHIOU}{Ohio University, Athens, Ohio  45701}
\newcommand*{\NOWISU}{Idaho State University, Pocatello, Idaho 83209}
\newcommand*{\NOWBRESCIA}{Universit\`{a} degli Studi di Brescia, 25123 Brescia, Italy}

\author{N.~Zachariou}
 \email{nicholas@jlab.org}
\affiliation{\YORK}

\author{E.~Munevar}
% \affiliation{Universidad Distrital Francisco José de Caldas}%
\affiliation{\UDFJC}

\author{B.L. Berman}\thanks{Deceased}
\affiliation{\GWUI}

\author{P.~Byd\v{z}ovsk\'{y}}
% \affiliation{Nuclear Physics Institute of the Czech Academy of Sciences, 250 68 \v{R}e\v{z}, Czechia}   % for all of us
\affiliation{\NPICA}

\author{A.~Ciepl\'{y}}
% \affiliation{Nuclear Physics Institute of the Czech Academy of Sciences, 250 68 \v{R}e\v{z}, Czechia}   % for all of us
\affiliation{\NPICA}

\author{G. Feldman}
\affiliation{\GWUI}

\author{Y. Ilieva}
\affiliation{\SCAROLINA}

\author{P. Nadel-Turonski}
% \affiliation{CFNS Stony Brook University, Stony Brook, NY 11794}%
\affiliation{\CFNS}

\author{D.~Skoupil}
% \affiliation{Nuclear Physics Institute of the Czech Academy of Sciences, 250 68 \v{R}e\v{z}, Czechia}   % for all of us
\affiliation{\NPICA}

\author{A.V.~Sarantsev}
% \affiliation{Helmholtz-Institut fuer Strahlen- und Kernphysik, Universit\"{a}t Bonn, 53115 Bonn, Germany}
\affiliation{\HELMBONN}

% ...

% \author {F.~Sabatie} 
%  \affiliation{\SACLAY}

\collaboration{CLAS Collaboration}%\noaffiliation

\date{\today}% It is always \today, today,
             %  but any date may be explicitly specified

\begin{abstract}
We report a new measurement of the beam-spin asymmetry \bsa{} for the \rctn{} reaction using quasi-free neutrons in a liquid-deuterium target. The new dataset includes data at previously unmeasured photon energy and angular ranges, thereby providing new constraints on partial wave analyses used to extract properties of the excited nucleon states. The experimental data were obtained using the CEBAF Large Acceptance Spectrometer (CLAS), housed in Hall B of the Thomas Jefferson National Accelerator Facility (JLab). The CLAS detector measured reaction products from a liquid-deuterium target produced by an energy-tagged, linearly polarised photon beam with energies in the range 1.1 to 2.3 GeV. Predictions from an isobar model indicate strong sensitivity to $N(1720)3/2^+$, $\Delta(1900)1/2^-$, and $N(1895)1/2^-$, with the latter being a state not considered in previous photoproduction analyses. When our data are incorporated in the fits of partial-wave analyses, one observes significant changes in $\gamma$-$n$ couplings of the resonances which have small branching ratios to the $\pi N$ channel.

%The high-precision data agree well with the sparse previous data. T MORE ON MODELS

\end{abstract}

\maketitle
%-------------------------------------------------  
\section{\label{sec:intro} Introduction}
The excitation spectrum of the nucleon provides fundamental information on the dynamics and interactions of its constituents, the quarks and gluons, and is an important tool to achieve a more detailed understanding of the nature of Quantum Chromodynamics (QCD) in the non-perturbative regime. Phenomenological constituent quark models~\cite{ref:SCapstick2000,ref:SCapstick1998, ref:SCapstick1985, ref:Loring2001, ref:Glozman1998,ref:Giannini2001} and lattice QCD~\cite{ref:JDudek1,ref:JDudek2,ref:JDudek3} predict a plethora of excited states of the nucleon that have yet to be experimentally determined. Alternative interpretations of nucleon structure that result in a reduced number of excited states (and therefore fewer ``missing" resonances) have also been proposed~\cite{RevModPhys.65.1199,Brodsky2007,KOLOMEITSEV2004243,Afonin20074537}. Experimentally establishing the existence, or absence,  of these missing nucleon resonances in nature has thus the potential to provide important insights into fundamental nucleon structure. As a result, the investigation continues to be a major focus at the world's leading electromagnetic beam facilities. Here we report a new precise measurement of the single polarisation observable, \bsa{}, and we discuss the effect this dataset has on partial wave analyses and models that aim at understanding the excited spectrum of nucleons.

The clean extraction of the nucleon excitation spectrum from experiment is complicated by the fact that the excited states are short-lived (broad) and overlapping. This complicates the extraction of their fundamental properties (photocouplings, lifetimes, spins, parities, decay branches, and even existence), with the difficulties exacerbated for states that produce weak signals in the decay channel under study. In the photoproduction of a pseudoscalar meson off the nucleon, the excited nucleon states contribute through their initial photoexcitation from the nucleon followed by the strong decay of the state. 
Values of the four complex amplitudes may be extracted up to an arbitrary phase, given data from a suitable combination of polarization measurements of sufficient accuracy, which would therefore provide a maximum constraint on subsequent partial-wave analyses~\cite{ref:WTChiang1997}.
%A ``complete experiment" for this reaction process can be obtained from measurement of a well chosen combination of eight observables. Such a complete experiment would enable unambiguous determination of the four complex reaction amplitudes in this photoproduction process and therefore a maximum constraint on subsequent partial-wave analysis~\cite{ref:WTChiang1997}. 
It has been recently argued that a reduced requirement on the number of measured observables may still allow convergence to a unique set of multipole amplitudes~\cite{PhysRevC.95.015206}. 

It is clear that eliminating the ambiguities in partial-wave analysis extraction of the excited nucleon states requires a precise and complete set of measurements of single- and double-polarisation observables, involving polarised beams, targets, and recoiling baryon polarimetry~\cite{ref:WTChiang1997,ref:JNyz2016,ref:AMSandorfi2011,ref:GFasano1992,ref:GKeaton1996}. Furthermore, measurements on both proton and (more challenging) neutron targets are indispensable, since resonances can have isospin dependent photocouplings~\cite{ref:AMSandorfi2013,ref:TMartBenholdHyde1995}. Additionally, the predicted differences in the preferred decay branches of individual states~\cite{ref:SCapstick1998,ref:WTChiang1997,ref:SCapstick19982}, mean that measurement of a wide range of pseudoscalar meson photoproduction final states, including $N\pi$, $K\Lambda$, $K\Sigma$ are crucial, and even data on vector meson ({\it e.g.,} $N\omega$) or multiple meson decays ({\it e.g.,} $N\pi\pi$) could be necessary. A recent review of the available results on non-strange baryon spectroscopy is given in Ref.~\cite{IRELAND2020103752}.

The relative importance of decay channels to strange quark containing particles ({\it e.g.,} $K\Lambda$, $K\Sigma$) for missing or poorly established states has been emphasized by constituent quark model calculations~\cite{ref:SCapstick1998}. Recent measurements of exclusive photoproduction of $K\Lambda$ and $K\Sigma$ from proton targets~\cite{PhysRevLett.119.062004,PhysRevC.93.065201} was key to achieve sensitivity to the newly discovered states reported in the PDG~2020~\cite{PDG2020}.  However, the corresponding data from neutron targets 
are much more limited. Although the differential cross section for $K^+\Sigma^-$~\cite{PhysRevLett.97.082003,PEREIRA2010289} and $K^0\Lambda$~\cite{PEREIRA2010289} reactions are measured with good precision, only one single polarisation measurement exists. The beam-spin asymmetry, \bsa{}, was originally obtained at LEPS~\cite{PhysRevLett.97.082003}, having kinematical coverage only at very forward kaon angles.
The few double-polarisation measurements, for $K^+\Sigma^-$~\cite{2020135662},  $K^0\Lambda$, and $K^0\Sigma^0$~\cite{PhysRevC.98.045205}, have more complete kinematic coverage but modest statistical accuracy, limiting definitive interpretations about contributing resonant states in partial wave analyses.  It was highlighted in the most recent work on the beam-target helicity asymmetry in $K^+\Sigma^-$ photoproduction~\cite{2020135662} that the \bsa{} at backward kaon angles showed an enhanced sensitivity to the contribution of the $N(2120){3/2^-}$ ($D_{13}$) excited state, which was found to improve the interpretation of the beam-target helicity asymmetry data~\cite{2020135662}.

In this work, we provide new experimental data on the beam-spin asymmetry \bsa{} for the reaction \rctn{}, for the first time covering a wide range of kinematics and in previously unexplored mass ranges for contributing states. The experiment used a linearly-polarised tagged-photon beam incident on a (bound) quasi-free neutron target (liquid deuterium). The paper is ordered as follows: Sec.~\ref{sec:setup} provides a brief description of the experimental setup,  Sec.~\ref{sec:asymmetry} describes the method we employed to determine the observable \bsa{}, and  details of the data analysis procedure and a description of the systematic uncertainties are discussed in Secs.~\ref{sec:analysis} and ~\ref{sec:Syst}, respectively. The results and a discussion of their implications are presented in Sec.~\ref{sec:Discussion}. 

%-------------------------------------------------  
\section{\label{sec:setup} Experimental setup}
The data for this work were collected during the E06-103 experiment~\cite{G13prop}, which was conducted at the Thomas Jefferson National Accelerator Facility (JLab) utilising the Continuous Electron Beam Accelerator Facility (CEBAF) and the CEBAF Large Acceptance Spectrometer (CLAS)~\cite{ref:BAMecking} housed in Hall B. The CLAS detector was comprised of a drift-chamber (DC) tracking system, a time-of-flight (ToF) system, and a calorimeter system that allowed particle identification and four-vector determination for charged and neutral particles. The charged  particles' momenta were determined by tracing them as they traversed a toroidal magnetic field, providing momentum resolution of $\sigma_p/p\sim1\%$. A start counter (ST)~\cite{SHARABIAN2006246} that surrounded the target cell provided the event start-time information in photoproduction experiments. The ST, in conjunction with the ToF system, was used to determine the speed of charged particles. The E06-103 experiment utilised  a 40-cm long liquid-deuterium target, centered 20~cm upstream of the nominal CLAS center to maximize acceptance for hyperon decays. Overall, the CLAS detector provided an efficient detection of charged particles over a large fraction of the full solid angle (between 8$^{\circ}$ and 142$^{\circ}$ in polar angles with $\sim$83\% azimuthal coverage). 

Hall B also housed the Tagger Facility~\cite{ref:DISober2000}, which enabled the selection and characterisation  of the photons that initiated the photo-induced reactions detected within the CLAS detector on an event-by-event basis. 
The real photon beam was produced via the bremsstrahlung technique, by impinging a monochromatic electron beam on a thin radiator. The post-bremsstrahlung electrons were momentum analysed in a magnetic spectrometer that provided energy and timing information of the incident photon beam. With an energy resolution of $\sim$0.2\%, this system permitted the tagging of photons with energies between 20\% and 95\% of the incident electron beam energy. The production of linearly polarized photons was based on the coherent bremsstrahlung radiation technique~\cite{ref:Uberall} utilising a 50-$\mu$m thick diamond radiator. With the use of a precise goniometer, data for two orientations of the photon polarisation were collected: one parallel ({\it Para}) and one perpendicular ({\it Perp}) to the Hall-B floor. Data were also obtained using an amorphous carbon radiator that enabled the determination of the degree of photon polarisation as discussed in the next section. For a fixed electron energy, the position of the coherent edge~\footnote{The coherent edge refers to the sharp falling edge in the enhancement spectrum as indicated in Fig.~\ref{Fig:PhotonPol}. } was selected by appropriately orienting the diamond radiator. Data were obtained for different electron beam energies, varying from 3.3 to 5.2 GeV, to enhance the degree of photon polarisation in six coherent peak positions, in steps of 200~MeV between 1.1 and 2.3~GeV.

\section{\label{sec:asymmetry} Beam-spin asymmetry}
The differential cross section for meson photoproduction off an unpolarised target with a linearly polarised photon  beam is given by~\cite{barker75}:
\begin{eqnarray}
\frac{d\sigma}{d\Omega}&=&\left(\frac{d\sigma}{d\Omega}\right)_0[1-P_{lin} \bsa \cos(2\eta)], \label{Eq:CSeq}
\end{eqnarray}
where $P_{lin}$ is the magnitude of the beam polarisation vector at an angle $\eta$ to the reaction plane\footnote{The reaction plane is defined by the cross product of the incoming photon and the outgoing meson.}. The above equation is obtained by integrating over the angular distribution of the hyperon decay products~\footnote{The full cross section equation as shown in Ref.~\cite{barker75} depends on additional double polarisation observables accessible by studying the angular dependence of the hyperon decay products. Integrating over the nucleon angle in the hyperon rest frame would eliminate such contributions only when the detector acceptance is uniform in these kinematics.  In principle, the detector acceptance might affect the implied integration over the angles of the hyperon decay products. However, as the self-analyticity of the $\Sigma^-$ is very small ($\alpha=0.068$), any double polarisation observables contributions/effects to the beam-spin asymmetry are negligible compared to the quoted systematics.} 
($\Sigma^-\to n\pi^-$ with a 99.85\% branching ratio). 
The determination of \bsa{} was done using a maximum likelihood approach. The likelihood function for a given event, $i$, taken from the cross-section Eq.~(\ref{Eq:CSeq}) is 
\begin{eqnarray}
L_i=c\left[ 1-P_{lin}^i\boldsymbol{\Sigma}\cos(2\eta_i)\right] A,
\end{eqnarray}
where $c$ is a normalisation coefficient and $A$ is the detector acceptance. In the construction of the log-likelihood function an approximation was made concerning the detector acceptance. Specifically, an acceptance that is largely independent of the kinematic variable $\eta$ was assumed, which resulted in a normalisation coefficient that is independent of the value of the polarisation observable. This approximation significantly simplified the extraction of the observable, but could potentially result in systematic biases. Extensive studies of such systematic effects showed that any residual effects on the polarisation observable are negligible.

The log-likelihood function that was maximized to obtain the polarisation observables is thus given by
\begin{eqnarray}
\log L&=& b+\sum_i \log\left[ 1-P_{lin}^i\boldsymbol{\Sigma}\cos(2\eta_i)\right] \label{Eq:LogLike},
\end{eqnarray}
where the constant $b$ is the observable-independent constant that absorbs the normalisation coefficient (associated with the photon flux) and the detector acceptance. The summation is over all events within a given kinematic bin. A transformation from the reaction frame (where the $y$ axis is perpendicular to the reaction plane)  to the lab frame (where the $y$ axis is vertical to the Hall B floor) was done using the following equations for the two orthogonal orientations of the photon polarisation ({\it Para} and {\it Perp})
\begin{eqnarray}
\eta^{Para}&=&-(\phi-\phi_0)\nonumber\\
\eta^{Perp}&=&\frac{\pi}{2}-(\phi-\phi_0),\nonumber
\end{eqnarray}
where $\phi$ is the meson azimuthal angle as measured in the lab frame, and $\phi_0$ is the offset of the photon polarisation with respect to the lab $x$ (for {\it Para}) or $y$ (for {\it Perp}) axis. Using the above two equations, Eq.~(\ref{Eq:LogLike}) can be written as
\begin{eqnarray}
\log L&=& b+\sum_i \log\left[ 1-\mathbb{P}_{lin}^i\boldsymbol{\Sigma}\cos(2\phi_i-2\phi_0)\right] \label{Eq:LogLike2},
\end{eqnarray}
where $\mathbb{P}_{lin}^i=P_{lin}^i$ for {\it{Para}} events, and $\mathbb{P}_{lin}^i=-P_{lin}^i$ for {\it{Perp}} events\footnote{The sign of $\mathbb{P}_{\gamma}$ absorbs the sign from the trigonometric function when translating $\eta$ to $\phi$, since $\cos(180-2\phi)=-\cos(2\phi)$.}. This likelihood function was maximized using MINUIT~\cite{minuit} to obtain the value of the observable \bsa{} and its uncertainty. The $\phi_0$ offset was determined  using a high-statistics channel from the same dataset (single pion photoproduction), found to be consistent with zero.

The determination of the beam-spin asymmetry requires a precise knowledge of the degree of photon polarisation, $P_{lin}^i$. The determination of $P_{lin}^i$ involved  using  the  coherent and incoherent bremsstrahlung  spectra  to  obtain an enhancement distribution that was then fit by a spectrum obtained from theoretical bremsstrahlung calculations (referred to as the Analytical Bremsstrahlung Calculation -- ANB). Specifically, the enhancement distribution was obtained by taking the ratio of the photon energy spectrum from the diamond radiator to one obtained using the amorphous radiator, and was used to constrain the relative contribution of the coherent and incoherent bremsstrahlung to the total photon yield. This ratio also removed Tagger channel efficiency fluctuations allowing a precise determination of the degree of photon polarisation. Subsequently, the enhancement plot was fit with the theoretical spectrum from ANB. More details on the procedure can be found in Refs.~\cite{ref:Uberall,ref:Livingston2011,ref:ZachariouThesis}.

\begin{figure}[htbp]
  \centering
      \includegraphics[width=0.47\textwidth]{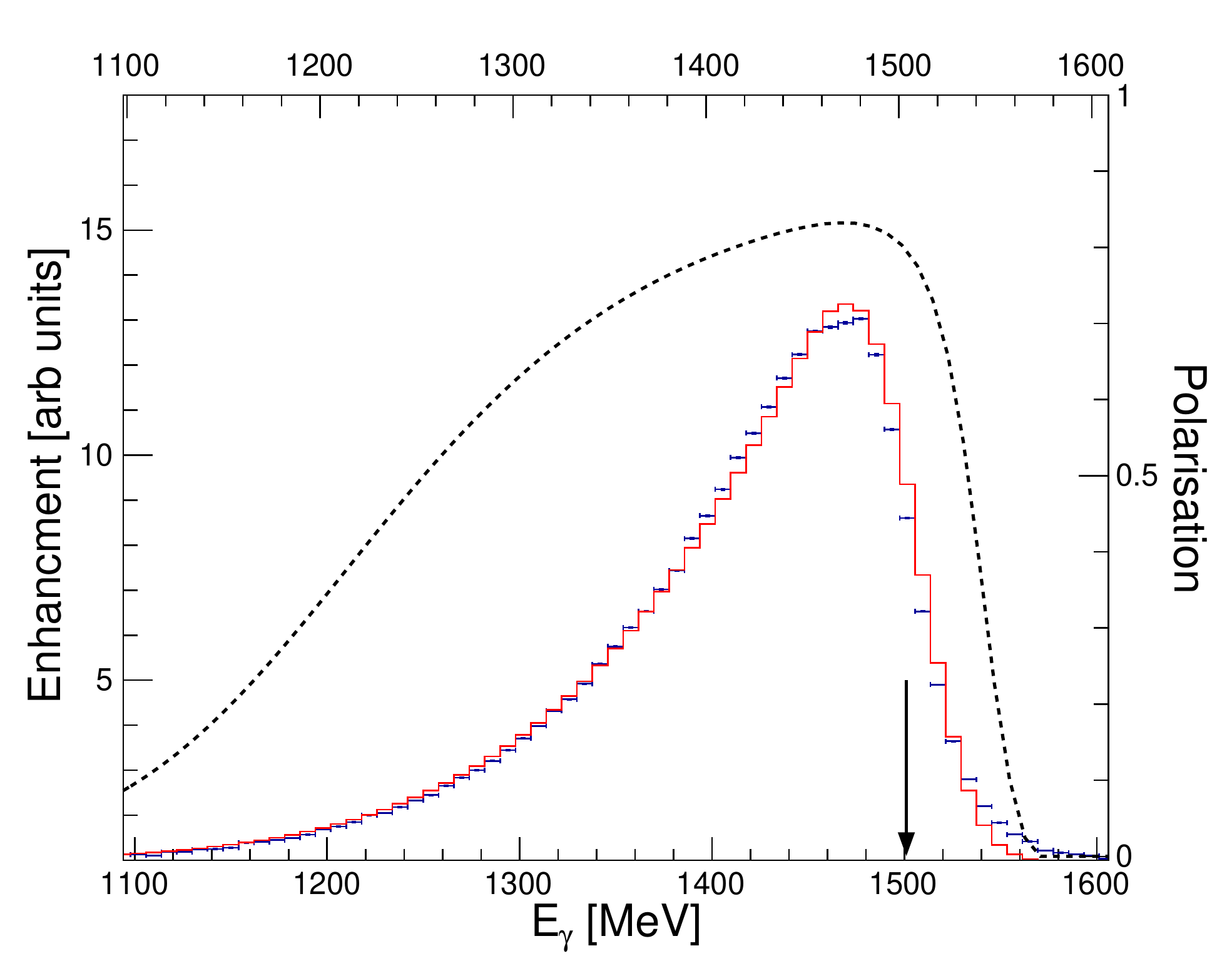}
  \caption{Example of an enhancement distribution (blue points) fit with the ANB calculation (red histogram) to determine the photon polarisation (dashed line). The arrow indicates the coherent edge position.}
  \label{Fig:PhotonPol}
\end{figure}

The ANB calculation takes into account 17 experimental parameters characterizing the geometry of the radiator, collimator, and photon beam. Several of these parameters were measured experimentally (such as the photon beam energy and beam spot size), whereas others (such as electron beam divergence on the radiator) were varied until a good agreement was obtained between the enhancement plot and the ANB calculation. These parameters were then used to calculate the degree of polarisation as a function of photon energy. An example of a fit to an enhancement spectrum with the ANB calculation is shown in Fig.~\ref{Fig:PhotonPol} along with the calculated photon polarisation (dashed line). This procedure was done for the various coherent-edge positions, allowing the determination of the photon polarisation on an event-by-event basis. The degree of photon polarisation throughout the experiment was on average 72\%.
%-------------------------------------------------  
\section{\label{sec:analysis} Data analysis}
The reaction of interest was reconstructed by selecting events with exactly one negative pion and one positive kaon, identifying the photon that initiated it, and applying the missing-mass technique under the assumption that the target was a nucleon at rest. Particle identification was done by following the standard procedures adopted for E06-103 analyses, by comparing the particle's speed calculated from two independent measurements: time-of-flight and momentum, with the latter requiring an assumption about the particle's rest mass.
The photon that initiated the reaction detected in CLAS was identified by timing coincidence at the event vertex between the tracks in CLAS and photons, with the latter being reconstructed using information from the Tagger spectrometer. The 2-ns beam bunch structure of the delivered electron beam allowed an unambiguous identification of the photon that initiated the reaction for $\sim90\%$ of the events. The remaining $10\%$ of events were associated with two or more photons with coincidence times within $\pm 1$~ns, and such events were discarded from further analysis. 
\begin{figure}[!h]
  \centering
     \includegraphics[width=8.6 cm]{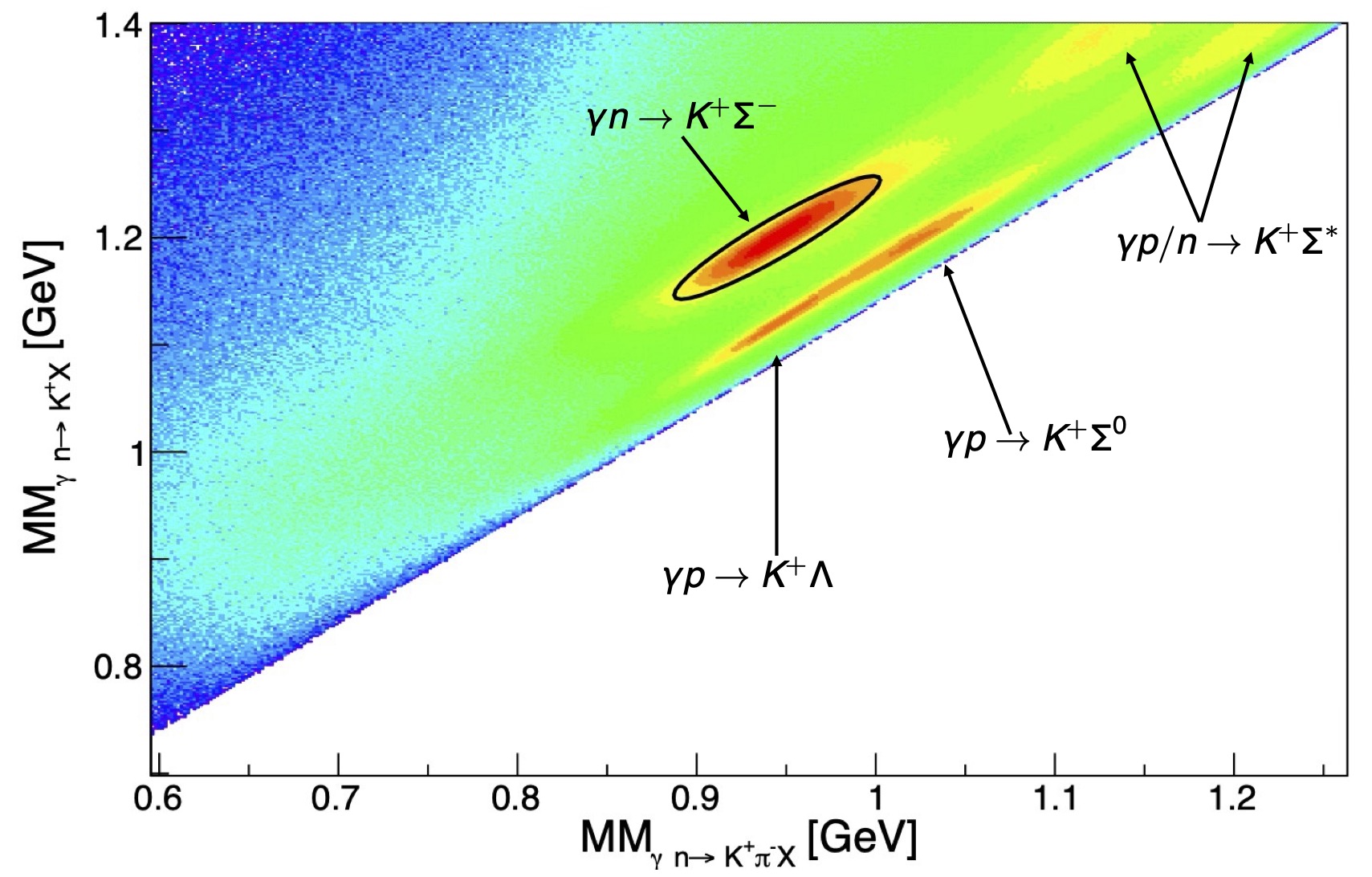}
  \caption{Mass of the missing state $X$ of $\gamma n\rightarrow K^+ X$ vs.  $\gamma n\rightarrow K^+\pi^-X$ indicating the different physics channels that contribute to the event sample.}
  \label{Fig:MMk_MMkpi}
\end{figure}

A fraction of positive pions from the reactions $\gamma N \to \pi^+\pi^- X$ (where $N$ can be either a proton or a neutron) were misidentified as kaons. Contributions from these events were eliminated by applying a cut on the mass of the missing state $X$ in $\gamma N \to \pi^+\pi^- X$ (assuming the reconstructed kaon was a misidentified pion).

 The reaction of interest was identified by further exploiting the missing-mass technique. Specifically, the correlation in the missing mass, $m_X$, distribution of $\gamma n \to K^+X$ ($MM_{\gamma n \to K^+X}$) and the  $m_X$ distribution  of $\gamma n\to K^+ \pi^-X$ ($MM_{\gamma n\to K^+ \pi^-X}$) allows a clean identification of the reaction of interest. This correlation is shown Fig.~\ref{Fig:MMk_MMkpi} along with the elliptical (two-dimensional) cut employed to select the events of interest. 
The parameters of the elliptical cut were optimised using simulated data (processed through a realistic detector simulation). This approach resulted in an event sample where background contributions were minimised, while retaining a large fraction of good events. With the parameters of the adopted cut, the average background contributions were found to be below 2\% (with such contributions accounted for in the  systematic  uncertainty,  as listed in Table~\ref{ref:TablSystUn}). 

%-------------------------------------------------  
\section{\label{sec:Syst} Systematic Uncertainties}
An extensive investigation of potential sources of systematic uncertainty was carried out with estimates summarised in Table~\ref{ref:TablSystUn}. Most sources have negligible contributions compared to the statistical uncertainty of the data.  The largest contribution originates from uncertainties of the degree of photon polarisation, and the second largest is due to the dilution of the measured observable stemming from having a bound rather than a free neutron target. 
\begin{figure*}[!ht]
\includegraphics[width=18.1 cm]{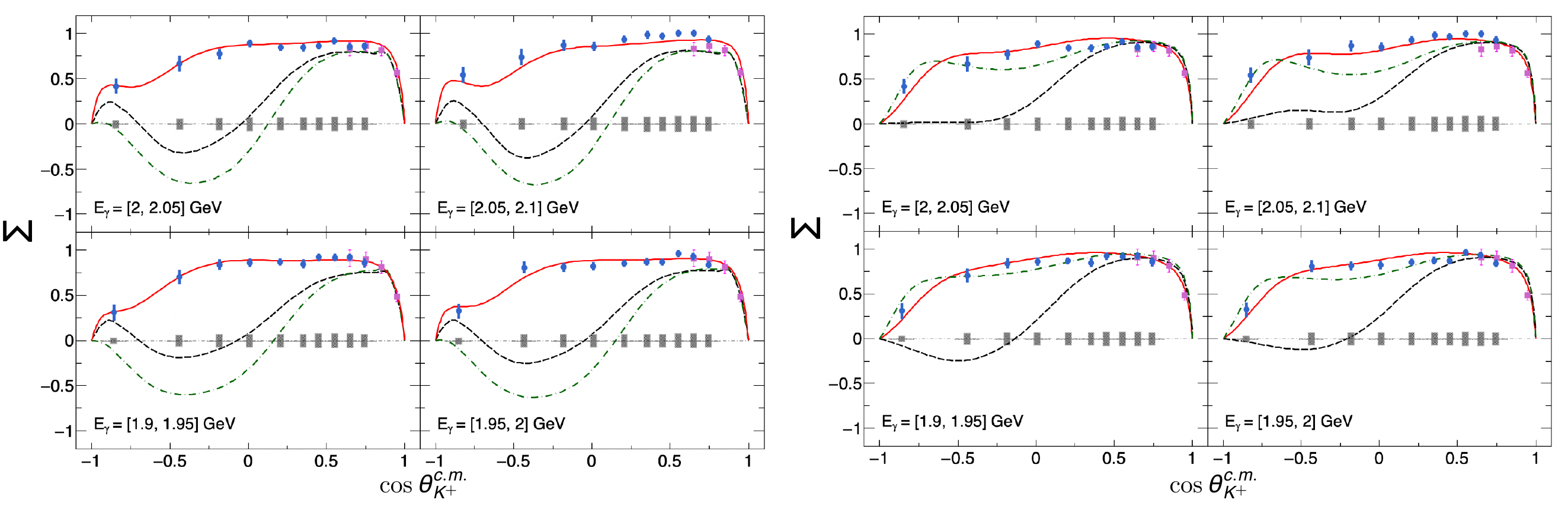}
\caption{Beam spin asymmetry \bsa{} as a function of kaon angle in the $c.m.$ for four photon energy bins as indicated in the panels. Experimental data from this work are shown with solid blue circles, whereas magenta points show the previously published results from LEPS~\cite{PhysRevLett.97.082003}. Statistical uncertainties are indicated with the errors bars, whereas the total systematic uncertainties of the CLAS results are shown by the shaded bar chart. The left set of panels shows the new Bonn-Gatchina solution that was fit to our data (red solid lines), as well as previously published Bonn Gatchina solutions with (green dash dotted lines) and without (black dashed lines) contributions from the $D_{13}$ resonance (see Ref.~\cite{2020135662} for a detailed discussion). The right set of panels show the full solution of the isobar model (red solid lines), as well as the solution without the $N(1720)\:3/2^{+}$ (black dashed lines), and the $\Delta(1900)\:1/2^{-}$ (green dashed dotted lines) resonance.\label{Fig:OldPredictions}}
\end{figure*}
The latter arises from effects of the Fermi motion of the target neutron and Final State Interactions  (FSI) of the outgoing reaction products with the deuterium remnants. Such dilution effects were investigated in detail using a smaller subset of the data sample in which the final-state neutron was detected in addition to the  $K^+$ and $\pi^-$~\cite{ref:MunevarThesis}, as well as through simulations. Only a weak dependence of \bsa{} on the momentum of the target neutron was discovered. The asymmetric (positive) systematic uncertainties reflect the fact that FSI effects only dilute and do not enhance the measured \bsa{}. Additional sources, including background contributions and misidentified kaon events, contributed to a much smaller degree as summarised in Table~\ref{ref:TablSystUn}.  
The uncertainties are split in two categories: an absolute uncertainty that is the same for all kinematics, and a relative uncertainty (associated with the photon polarisation) with its magnitude determined for each point. 
% Further details on the systematic uncertainties can be found in the online supplementary documentation.

\begin{table}[!ht]
\centering
\begin{tabular}{|c   c|}
\hline 
Source & $\sigma^{sys}$\\\hline
\hline\hline
Maximum Likelihood  &negligible \\
Kaon PID & $\pm 0.008$ \\
Pion PID & $\pm 0.004$ \\
Photon selection & $\pm0.002$ \\
Misidentified kaons& $\pm0.003$ \\
Kaon Decay in flight & $\pm0.0064$ \\
$\Sigma^*$ background contribution & $\pm0.007$ \\
$\Lambda$ and $\Sigma^0$ contributions & $\pm0.008$ \\
Fiducial cut & $\pm0.002$ \\
FSI & +0.024 \\
\hline
\bf{Total Absolute Systematic} & $\mathbf{^{+0.029}_{-0.016}}$\\
\hline
\bf{Photon polarisation} &\bf{8\%} \\
% \hline
% \bf{Total Scale Systematic} & \bf{8.4\%+}\\
 \hline\hline
\end{tabular}
\caption{Summary of systematic uncertainties  of \bsa. \label{ref:TablSystUn} }
\end{table}

%-------------------------------------------------  

\section{\label{sec:Discussion} Results and Discussion}
\begin{figure*}[!ht]
\includegraphics[width=18.1 cm]{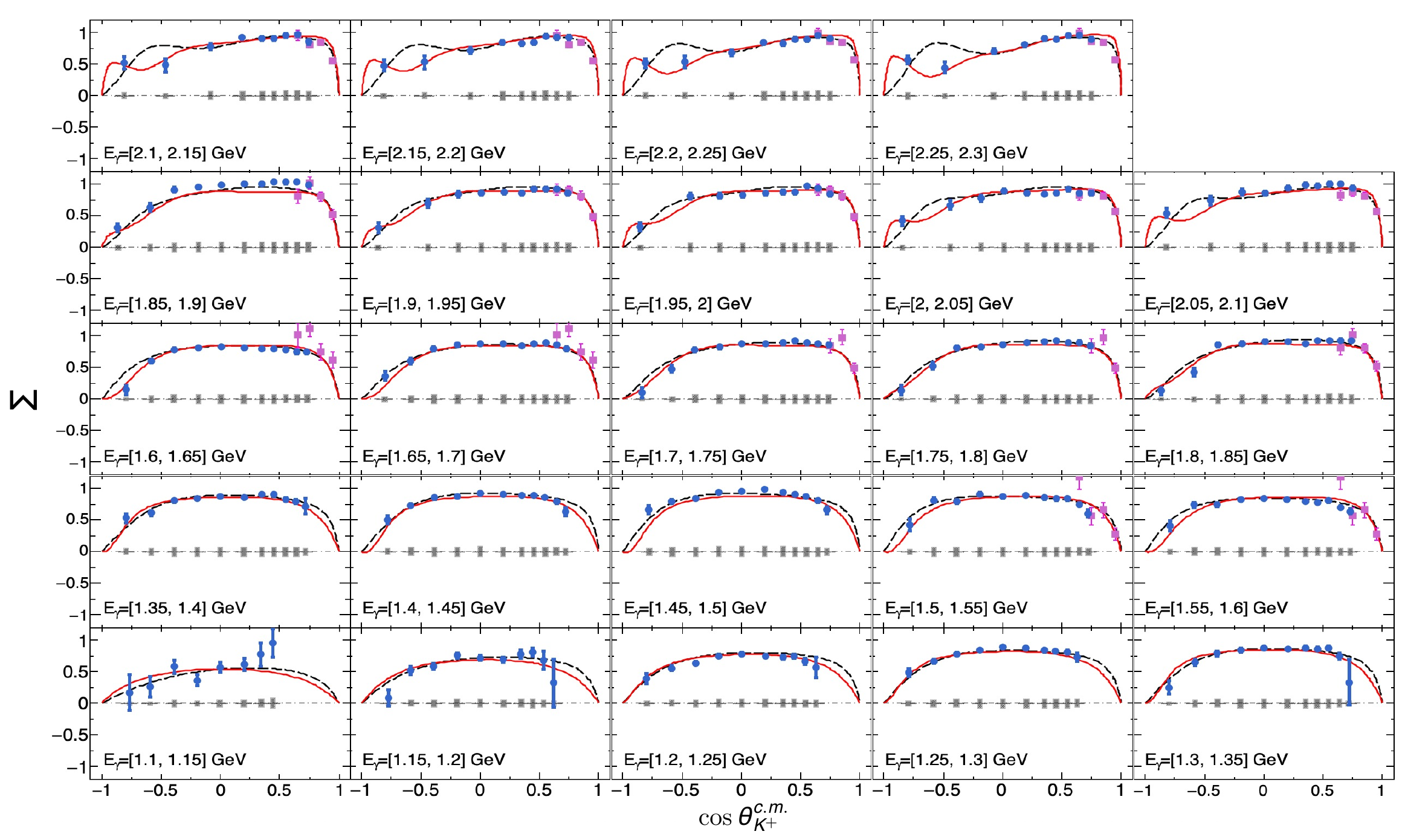}% Here is how to import EPS art
\caption{\label{Fig:SigmaResults}Beam spin asymmetry \bsa{} as a function of kaon angle in the $c.m.$ The different panels show bins in photon energy. Experimental data from this work are shown with solid blue circles, whereas magenta points show the previously published results from LEPS~\cite{PhysRevLett.97.082003}. Statistical uncertainties are indicated with the errors bars, whereas the total systematic uncertainties of the CLAS results are shown by the shaded bar chart. The black dashed line indicates the full solution of the isobar model as described in the text and the red solid line indicates the new Bonn-Gatchina solution that was fit to our data.} 
\end{figure*}

The extracted \bsa{} data are shown by the blue solid circles in Figs.~\ref{Fig:OldPredictions} and~\ref{Fig:SigmaResults}, binned in 50-MeV wide photon-energy bins (from 1.1 to 2.3~GeV) and in 10 bins of kaon production angle in the center-of-momentum ($c.m.$) frame\footnote{Commonly known as the center-of-mass frame.}, $\cos\theta_{K^+}$. Figure~\ref{Fig:OldPredictions} shows how the new precise results in four photon-energy bins compare with previous and current Bonn Gatchina solutions (left) or with an isobar model predictions that focuses on contributions from specific resonance states (right). Figure~\ref{Fig:SigmaResults} compares the two new solutions for all available kinematic bins as discussed in detailed later on.  
The angular bins are  contiguous  but with  varying  widths to accommodate the angular variation of the reaction yield as to keep the statistics per bin rather constant. 
The statistical uncertainties are shown by the error bars for each point, and the systematic uncertainties are shown by the grey bands. For all photon energies, the measured \bsa{} is large, positive, and for forward-central kaon angles rather uniform.  The data exhibit a fall off at backward kaon angles, with  \bsa{} typically larger in the forward angle region.  As \bsa{} must have a value of 0 at  $\cos\theta_{K^+}=\pm1$,  the observable  values  outside  of  our  acceptance  range  ({\it i.e.,} between $\cos\theta_{K^+}=0.75$ and $\cos\theta_{K^+}=1.0$ for forward angles)  must vary rapidly to reach 0. At backward angles the transition to zero exhibits a more gradual trend. 

The published results for \bsa{} from LEPS~\cite{PhysRevLett.97.082003} are also shown by the magenta solid squares in Figs.~\ref{Fig:OldPredictions} and~\ref{Fig:SigmaResults}. The LEPS data are limited to very forward kaon angles and have larger statistical and systematic uncertainties than our data. Nevertheless, the results from CLAS are in good agreement with these previously published data (note the LEPS data were obtained in 100-MeV wide photon energy bins). The improvement in the quality and range of available data with this new measurement is apparent in Fig.~\ref{Fig:SigmaResults}.

% The most recent predictions from the Bonn-Gatchina model ~\cite{2020135662} with (without) the proposed $D_{13}$ resonance are shown by the dash-dotted magenta (long-dashed cyan) lines. These predictions are fit to the current world dataset in meson photoproduction, excluding the data from this work.  It is clear that neither of these solutions can  reproduce the angular dependence of \bsa{}, with discrepancies especially apparent at photon energies above 1.3~GeV. Clearly, the new data have the potential to impact this partial-wave analysis and, therefore, the excited nucleon spectrum therein.

The solutions of the Bonn-Gatchina group BG2016~\cite{2020135323} (not shown here) predicted the beam
asymmetry for the $\gamma n\to K^+\Sigma^-$ reaction above 1850 MeV
to be negative at the backward and the central angular region. 
The most recent predictions from the Bonn-Gatchina model ~\cite{2020135662} with (without) the proposed $D_{13}$ resonance are shown by the dash-dotted green (dashed black) lines in the left panels of  Fig.~\ref{Fig:OldPredictions} (only four representative photon energy bins are shown). These predictions were fit to the current world dataset in meson photoproduction, including the unpolarised differential cross section, the data on the beam-target helicity asymmetry
measured by the CLAS Collaboration, and the LEPS data on the beam
asymmetry, but excluding data from this work. The \bsa{} data from LEPS were measured only in the very forward
angular region and mostly were defined by the contribution from the
$t$-channel exchange amplitudes. Moreover, the CLAS data on the
unpolarized cross section and beam-target helicity asymmetry did not cover the
very backward angular region, which allowed ambiguous solutions.  It is clear that neither of these solutions can  reproduce the angular dependence of \bsa{}, with discrepancies especially apparent at photon energies above 1.3~GeV.
% The new beam asymmetry data presented in this paper thus have a significant impact in this partial-wave analysis and, therefore, the excited nucleon spectrum therein.  
Clearly, the new data have the potential to impact this partial-wave analysis and, therefore, the excited nucleon spectrum therein.

% TheBonn-Gatchina analysis was based on the data on the unpolarized
% differential cross section and the data on the helicity asymmetry
% measured by the CLAS collaboration and the LEPS data on the beam
% asymmetry. The latter ones were measured only in the very forward
% angular region and mostly were defined by the contribution from the
% $t$-channel exchange amplitudes.  

The inclusion of the new beam asymmetry data
in the full combined analysis led to significant changes in the $\gamma$-$n$ couplings of the
resonances that have small branching ratios to the $\pi N$ channel.
The largest changes were found in the $D_{13}$ and $P_{13}$ partial
waves: here the states in the region above 1850 MeV were mostly seen
in the reactions with open strangeness. The newly obtained solution is indicated by the red solid curves in Figs.~\ref{Fig:OldPredictions} and~\ref{Fig:SigmaResults}. The detailed and systematic
analysis of this solution will be presented in a separate paper,
which will follow the present publication.

The full predictions from an isobar model~\cite{PhysRevC.93.025204} for \bsa{} are shown by the black dashed line in Fig.~\ref{Fig:SigmaResults}. These are based on an effective Lagrangian 
in a tree-level approximation. The non-resonant part of the amplitude consists of the Born terms and
exchanges of resonances in the $t$- ($K^*$ and $K_1$) and $u$-channels ($\Sigma^{*}$).
The main coupling constant $g_{K^+\Sigma^- n} = \sqrt{2}g_{K^+\Sigma^0 p}=1.568$, 
which determines the strength of the Born terms, was taken from 
the $K^{+}\Lambda$ channel \cite{PhysRevC.93.025204} and kept unchanged in the
present fit. The resonant part is modeled by $s$-channel exchanges of nucleon
and $\Delta$ resonances with masses below about 2~GeV. 
Hadronic form factors included in the strong vertices account for hadron
structure and regularize the amplitude at large energies. The form factors
are introduced in the way that keeps gauge invariance intact, in analogy with the
method used in Refs.~\cite{PhysRevC.93.025204} and \cite{PhysRevC.97.025202}. 
The solution presented in Fig.~\ref{Fig:SigmaResults} was fit to the current CLAS (and LEPS) \bsa{} data, as well as the differential cross section of $\gamma n \to K^+\Sigma^-$ from CLAS~\cite{PEREIRA2010289}. In total, 24 free parameters (22 couplings and 2 hadron form factor cut-offs) 
were used to fit 332 cross section data points and 284 asymmetries, all of them are restricted to energies up to $E_\gamma = 2.6$ GeV. The fit parameters of the isobar model were extracted adopting the procedure outlined in  Refs.~\cite{PhysRevC.93.025204,PhysRevC.97.025202} for the $K^{+}\Lambda$ channel. 
More details are provided in Ref.~\cite{Bydzovsky2021}.

The considered set of nucleon resonances in the isobar model was motivated by previous analyses of $K^{+}\Lambda$~\cite{PhysRevC.93.025204, PhysRevC.97.025202} and $K\Sigma$ photoproduction~\cite{PhysRevC.53.2613}. 
\begin{table}[h]
\caption{Characteristics of included resonances with their masses and widths taken 
as the PDG Breit-Wigner averages. The available branching ratios to the $K\Lambda$
and $K\Sigma$ channels are also taken from the PDG~\cite{PDG2020}. For the nucleon
and Delta resonances, the values $g_1$ and $g_2$ show the baryon-$K\Sigma$ scalar 
and tensor couplings obtained in our fit, while for the $K^{*}$ and $K_1$ states 
they represent the vector and tensor couplings, respectively. \label{tab:BCSfit}
}
\begin{center}
{\footnotesize{
\begin{tabular}{ccccccc}
\hline
\hline
Resonance      &  Mass    &  Width  & \multicolumn{2}{c}{Branching ratios}
                                    & \multicolumn{2}{c}{Couplings} \\ 
               & (MeV)    & (MeV)   & $\Lambda K$ & $\Sigma K$ &  $g_1$  &  $g_2$ \\
\hline \hline
$N(1535)\;1/2^-$ & 1530     & 150     &     ---     &    ---     & -0.709  &   ---  \\
$N(1650)\;1/2^-$ & 1650     & 125     &     0.07    &    0.00    &  0.314  &   ---  \\
$N(1675)\;5/2^-$ & 1675     & 145     &     ---     &    ---     & -0.013  &  0.022 \\
$N(1710)\;1/2^+$ & 1710     & 140     &     0.15    &    0.01    & -0.940  &   ---  \\
$N(1720)\;3/2^+$ & 1720     & 250     &     0.05    &    0.00    & -0.098  & -0.082 \\
$N(1875)\;3/2^-$ & 1875     & 200     &     0.01    &    0.01    & -0.220  & -0.223 \\
$N(1880)\;1/2^+$ & 1880     & 300     &     0.16    &    0.14    & -0.050  &   ---  \\
$N(1895)\;1/2^-$ & 1895     & 120     &     0.18    &    0.13    & -0.063  &   ---  \\
$N(1900)\;3/2^+$ & 1920     & 200     &     0.11    &    0.05    & -0.051  & -0.004 \\
$N(2060)\;5/2^-$ & 2100     & 400     &     0.01    &    0.03    & -0.00001 & 0.003 \\
$N(2120)\;3/2^-$ & 2120     & 300     &     ---     &    ---     & -0.034  & -0.010 \\ 
\hline
$\Delta(1900)\;1/2^-$ & 1860 & 250    &     ---     &    0.01    &  0.298  &   ---  \\ 
\hline
$K^*(892)$     & 891.7    & 50.8    &     ---     &    ---     &  0.366  &  1.103 \\ 
$K_1(1270)$    & 1270     & 90      &     ---     &    ---     & -1.448  &  0.473 \\ 
\hline
\end{tabular}
}}\end{center}
\end{table}
Some additional $N^{*}$ resonances predicted to strongly couple the $K\Sigma$ channel 
were also investigated in the analysis.
The variant with the smallest $\chi^{2}/ndf$ and reasonable values of the parameters 
was selected. The complete set of resonances from this best fit is provided in Table~\ref{tab:BCSfit} \footnote{Note that only the statistical uncertainties of the fit data were used in the computation of the $\chi^2$, which 
results in a relatively large value $\chi^{2}/ndf = 2.39$  for the selected solution. This approach was chosen due to missing systematic uncertainties in some data sets. When systematics are taken into account, the $\chi^2$ value typically drops without changing the quality of results.}. The solution indicates contribution from two kaon resonances, multiple nucleon resonances, one $\Delta$ resonance, and no hyperon resonances \footnote{The obtained couplings $g_1$ and $g_2$ from the fit 
listed in Table~\ref{tab:BCSfit} are all reasonable, as are the hadronic form factor cut-offs $\Lambda_{\rm bgr}=0.874$ GeV and $\Lambda_{N}=1.451$ GeV (see Ref.~\cite{PhysRevC.97.025202} for a description of these parameters).}. The combined asymmetry and cross section data show a strong sensitivity to $N(1720)\:3/2^{+}$, whose omission significantly diminishes both observables (see right panels of Fig.~\ref{Fig:OldPredictions}). Sensitivity was also observed from the $\Delta(1900)1/2^-$ resonance, specifically at central angles, as indicated by the green dashed dotted lines in the right panels of Fig.~\ref{Fig:OldPredictions}.
A significant contribution of the $N(1895)\:1/2^-$ state was obtained, a state not considered 
in previous photoproduction analyses, but with a relatively large $K\Sigma$ branching ratio. The role of hyperon resonances appears small, giving negligible effects on the predicted observables.

\section{\label{sec:Summary} Summary}
We present the first precise measurement of the beam-spin asymmetry  \bsa{}, employing a linearly polarised photon beam, for \rctn{} up to photon energies of $E_\gamma=2.3$~GeV. The new data obtained using a deuterium target agree well with previously published data from LEPS (limited only to forward angles), while significantly extending the available kinematic coverage for \rctn{} down to photon energies of $E_\gamma=1.1$~GeV and cover a large angular range.
The new \bsa{} data are an important addition to the world database and have a large effect on the determined $\gamma$-$n$ couplings of resonances that have small branching ratios to the $\pi N$ channels. The largest changes were found in the $D_{13}$ and $P_{13}$ partial
waves. A more detailed analysis in the Bonn-Gatchina framework will be presented in a planned joint publication. The new data were also fit using an isobar model based on an effective Lagrangian in a tree-level approximation, with results indicating contributions from two kaon resonances, multiple nucleon resonances (with significant contributions from the $N(1720)3/2^+$ and $N(1895)1/2^-$), one $\Delta$ resonance, and no hyperon resonances. Details on the isobar model will also be presented in a longer planned joint publication.

%-------------------------------------------------  
\section{\label{sec:Acknowledgments} Acknowledgments}
This work has been supported by the U.K. Science and Technology Facilities Council (ST/P004385/2, ST/T002077/1, and ST/L00478X/2) grants, as well as by the Czech Science Foundation GACR grant 19-19640S. We also acknowledge the outstanding efforts of the staff of the Accelerator and Physics Divisions at Jefferson Lab that made this experiment possible. The Southeastern Universities Research Association (SURA) operated the Thomas Jefferson National Accelerator Facility for the United States Department of Energy under contract DE-AC05-06OR23177. Further support was provided by the National Science Foundation, the Italian Istituto Nazionale di Fisica Nucleare, the Chilean Comisi\'on Nacional de Investigaci\'on Cient\'ifica y Tecnol\'ogica (CONICYT), the French Centre National de la Recherche Scientifique, the French Commissariat \`{a} l'Energie Atomique, and the National Research Foundation of Korea. 

\bibliography{biblio}

\end{document}